\documentclass[journal, twocolumn, 10pt]{IEEEtran}
\usepackage{graphicx}
\usepackage{algorithm}
\usepackage{algorithmicx}
\usepackage{algpseudocode}
\usepackage{cite}
\usepackage{color}
\usepackage{amsmath,amssymb,amsfonts,amsthm,dsfont}
\usepackage{bm}
\usepackage{setspace}
\usepackage{multirow}
\usepackage{mathtools}
\usepackage{epstopdf}
\usepackage[font+=small]{caption}
\usepackage[font+=small]{subcaption}

\begin{document}
\title{Decapitation via Digital Epidemics:  \\ A Bio-Inspired Transmissive Attack}
\author{
    \IEEEauthorblockN{Pin-Yu Chen\IEEEauthorrefmark{1}, Ching-Chao Lin\IEEEauthorrefmark{2}, Shin-Ming Cheng\IEEEauthorrefmark{2}, Hsu-Chun Hsiao\IEEEauthorrefmark{3}, and Chun-Ying Huang\IEEEauthorrefmark{4} 
 } \\
    \IEEEauthorblockA{\IEEEauthorrefmark{1}Department of Electrical Engineering and Computer Science, University of Michigan, Ann Arbor, USA
    \\ pinyu@umich.edu} \\
 \IEEEauthorblockA{\IEEEauthorrefmark{2}Department of Computer Science and Information Engineering, National Taiwan University of Science and Technology, Taipei, Taiwan
 	\\ \{m10415008, smcheng\}@mail.ntust.edu.tw} \\
\IEEEauthorblockA{\IEEEauthorrefmark{3}Department of Computer Science and Information Engineering, National Taiwan University, Taipei, Taiwan
\\ hchsiao@csie.ntu.edu.tw} \\
\IEEEauthorblockA{\IEEEauthorrefmark{4}Departement of Computer Science, National Chiao Tung University, Hsinchu, Taiwan
\\ chuang@cs.nctu.edu.tw}
}

\maketitle
\setstretch{1.0}
\thispagestyle{empty}
\begin{abstract}
The evolution of communication technology and the proliferation of electronic devices have rendered adversaries powerful means for targeted attacks via all sorts of accessible resources. In particular, owing to the intrinsic interdependency and ubiquitous connectivity of modern communication systems, adversaries can devise malware that propagates through intermediate hosts to approach the target, which we refer to as transmissive attacks.  Inspired by biology, the transmission pattern of such an attack in the digital space much resembles the spread of an epidemic in real life. This paper elaborates transmissive attacks, summarizes the utility of epidemic models in communication systems, and draws connections between transmissive attacks and epidemic models. Simulations, experiments, and ongoing research challenges on transmissive attacks are also addressed.

\end{abstract}
%\cite{Albert00}

%\IEEEpeerreviewmaketitle

\section{Introduction}
\label{sec_intro}
In recent years, researchers have successfully borrowed several biological mechanisms  from the nature for devising efficient protocols and understanding their performance via the associated mathematical models, especially for cyber security in communication systems \cite{Mazurczyk15,Cheng16}. Inspired by epidemiology, this paper investigates an emerging attack pattern, \textit{transmissive attack}, featuring heterogeneous propagation paths and specific targets. Analogous to the spread of epidemics in the nature, malicious codes act as viruses that are capable of infecting hosts (i.e., electronic devices) via various communication resources, and they can be stealthily transported by intermediate hosts to reach the primary hosts (i.e., targets), which is similar to the biological mechanism known as \textit{host specificity}.

Inevitably, the proliferation of electronic devices equipped with communication capabilities, and the  penetration  of Internet of Things have created ever-increasing security threats that we call as \emph{digital epidemics}, which may be even more vital than actual transmissive diseases like Dengue Fever, Ebola or SARS due to their cyber transmission and dormant operation nature, and their induced loss in properties and privacy. It is worth mentioning that although the fragility of modern communication systems may seem to be a shocking news to the world, the severe consequences caused by digital epidemics have been foreseen by researchers \cite{Wang09,CPY10,Cheng11}. In the past two decades various advanced communication technologies such as cellular systems and wired and wireless networks, and tremendous user activities such as online social networking and mobile applications have constituted a heterogeneous yet ubiquitous network among users and devices around the globe, which is known as a complex communication network \cite{Cheng13} or a generalized social network \cite{Cheng11}. Malicious codes are able to exploit these heterogeneous communication paths and intrinsic system interconnectivity for propagation and thereby compromise more devices.

By investigating recently discovered attack cases and system vulnerabilities, we present an emerging attack pattern named \emph{transmissive attack}, where an adversary can leverage diverse communication paths and common communication protocols (e.g., Internet of Things) to indirectly compromise a target (or a set of targets) that cannot be directly accessed by the adversary.
Furthermore, in order to increase the possibility of reaching the target, transmissive attacks may camouflage their activities to elude detections rather than indiscriminately infesting as many hosts as possible. The specificity to targeted attacks and heterogeneity in propagation paths distinct transmissive attacks from well-known Internet worms such as Code Red that only exploits single propagation resource (the Internet) and features indiscriminate attacks.

\begin{figure*}[]
	\centering
	\includegraphics[width=4.8in]{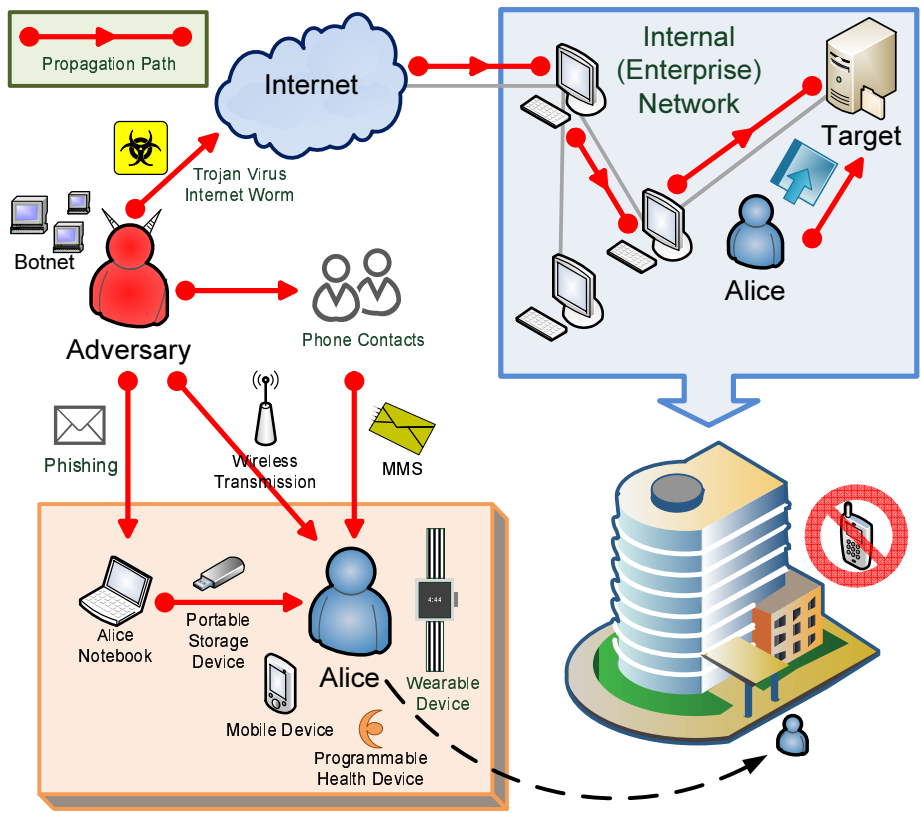}
	\caption{Illustration of transmissive attacks and their propagation paths. Transmissive attacks exploit various communication resources for propagation in order to reach the target. This diagram shows some examples of propagation paths that lead to the target.}
	\label{Fig_Transmissive_Attack}
\end{figure*}

Fig. \ref{Fig_Transmissive_Attack} illustrates several possible means for the adversary to access the target\footnote{A detailed attack scenario is provided in the supplementary file.}. Consider the target to be a personal computer in an enterprise that is granted access to private employee/customer databases or confidential corporate files, and all external networking connections to the target are prohibited. If the target is connected to other internal machines that connect to the outside world, the adversary can eventually reach the target by successive propagation. Pessimistically, even if all connections from other internal machines to the target are prohibited, the adversary can still manage to approach the target by compromising the authorized user's electronic devices such as portable storage devices, wearable devices or health devices embedded in human body equipped with communication capabilities. In practice all the adversary needs to launch a transmissive attack is simply release a malicious code (e.g., a Trojan virus) and then sits back and waits for the code to propagate among hosts (potential victims), either via cyber connection (e.g., phishing from the Internet) or human carrier (e.g., Bluetooth or WiFi direct reception from proximity), to create an indirect (i.e., multiple-hop) communication path for accessing the target. Moreover, after successful intrusion the adversary can erase its traces from the communication path (e.g., implementing a global timer for self-deactivation) to reduce the risk of being uncovered.

Inspired by biology, we use epidemic models to evaluate the consequences of a transmissive attack from a macroscopic system-level perspective. Analogous to disease transmission assessment, epidemic models categorize the hosts in a system into a few states to analyze the collective behavior of a system with parametric mathematical models (e.g., coupled state difference equations or Markov chains) for the purposes of status tracking, outbreak prediction, and further actions. As a first step toward analyzing transmissive attacks, we use epidemic models to investigate the probability of successfully compromising the target and quantify the risk of exposure with respect to time. We show that the tradeoffs in time between the probability of successful intrusion to the target and the associated risk can be characterized by epidemic models, thereby enabling security analysis\footnote{Although in recent years epidemic-like information propagation has been well studied in communications society in the contexts of ``epidemic routing'', where packets are transmitted in a store-and-forward fashion in intermittently connected networks, little is known on how to apply these well-developed analysis tools \cite{Haas06,Zhang07,CPY14control,peng2014smartphone} to model transmissive attacks and beyond.
	}.

%This paper is organized as follows. The following section provides an overview of epidemic models on information dissemination and malware propagation. Then we elaborate on how epidemic models can be used to evaluate the performance of transmissive attacks. For demonstration we conduct simulations and experiments of transmissive attacks in mobile networks and mobile social networks, respectively. Ongoing challenges and open research questions are also discussed.

%%%% XXX: begin new writing
\section{Transmissive Attacks in Practice}
As illustrated in Fig. \ref{Fig_Transmissive_Attack},
one typical scenario of transmissive attack is that an adversary aims to approach a target by propagating through intermediate hosts via all possible communication resources  in a complex communication system. The purpose of such an indirect propagation can be that direct access from the adversary to the target is unavailable, or the adversary attempts to hide his/her true identity by manipulating compromised machines to launch an attack, such as the exploitation of mobile devices as botnets \cite{Mtibaa15}. It is worth noting that a transmissive attack can be more insidious due to inherent configurability of electronic devices carried by a user, e.g., programmable in-body health devices or wearable mobile devices, which enables malware propagation even when typical communication devices such as cell phones and laptops are prohibited.

In addition to hidden identity, another appealing advantage of the transmissive attack is that the adversary need not to know the complete network topology to accomplish the attack. All the adversary needs to do is release a transmissive malicious code and then waits for the malicious code propagating to the target due to its transmissive nature. In practice a transmissive attack can be simply accomplished by devising a Trojan virus designed to be operated in the stealthy transmissive mode during propagation and activated when reaching the target. Advanced transmissive attacks can camouflage normal user/network activities to elude intrusion detection or system monitoring, thereby incurring severe threats to security and privacy. 

% XXX: merge HC's improvements
One of the most notable targeted attacks is the Stuxnet attack discovered in 2010. Stuxnet is designed to target a specific version of industrial control systems in a surreptitious manner, whereas traditional worms often aim to infest as many hosts as possible in a short time period. Stuxnet thus exhibits several distinguishing characteristics compared with traditional worms: each Stuxnet worm only replicates itself for at most three times; it is programmed to self-destruct on a day in 2012; it can stealthy propagate via carriers (i.e., vulnerable Windows computers) without showing any symptoms and only unpack its malicious payload when reaching a target; multiple zero-day vulnerabilities are used. Although performing a targeted attack may be expensive, and indeed Stuxnet is believed to be state-sponsored malware due to its unprecedented level of sophistication, more and more Stuxnet successors (e.g., dudu and Flame) demonstrate how far an adversary is willing to go for high-value targets.

\begin{table}
\begin{center}
\caption{Statistics of Vulnerabilities Identified on Popular Applications and Platforms.}
\label{tbl:vulnerability}
\begin{tabular}{|rccc|}
\hline
\hline
	& \multicolumn{3}{c|}{Number of Vulnerabilities} \\
Applications \& Platforms & 2013 & 2014 & 2015 \\
\hline
\hline
Adobe Acrobat Reader 	& 66	& 44	& 129	\\
Apple iPhone OS 	& 90	& 120	& 375	\\
Apple Mac OS X		& 65	& 135	& 384	\\
Apple WatchOS		& -	& -	& 53	\\
Google Android		& 7	& 11	& 130	\\
Microsoft Internet Explorer & 129	& 243	& 231 \\
Microsoft Office	& 17	& 10	& 40	\\
Microsoft Windows 7	& 100	& 36	& 147	\\
Linux Kernel		& 189	& 133	& 77	\\
\hline
\hline
\end{tabular}
\end{center}
\vspace*{0.02cm}
Source: http://www.cvedetails.com/
\end{table}

Stuxnet is one kind of Advanced Persistent Threat (APT), which can be seen as one specific case of transmissive attacks. An APT possesses the feature of specificity in targets. Although the feature of heterogeneity in community paths is not mandatory for an APT, it would shorten the process to approach the targets if heterogeneous community paths are considered. In 2013, Mandiant\footnote{APT1: Exposing One of China's Cyber Espionage Units, http://intelreport.mandiant.com/} summarized the attack life cycle of APT: 1)~Initial compromise, 2)~Establish foothold, 3)~Escalate privileges, 4)~Internal reconnaissance, 5)~Move laterally, 6)~Maintain presence, and 7)~Complete mission. An APT attack often loops through steps 3 to 6 until it reaches the specific target. These steps are also applicable to transmissive attacks.

To launch a successful transmissive attack, an attacker would also like to increase heterogeneity in community paths, e.g., by exploiting diverse vulnerabilities. The statistics of recently reported vulnerabilities (Table~\ref{tbl:vulnerability}) shows the numbers are consistently increasing for most platforms and applications, even for modern mobile and wearable devices. Consequently, various activities and mediums including web downloads, document reading, e-mail reading, short messages delivery, Wi-Fi access, Bluetooth access, and NFC contacts can be used together to deliver malicious payloads and approach the targets. By leveraging these existing vulnerabilities, an attacker can even launch transmissive attacks in the background and be invisible to a user.

For example, in July 2015, an unprecedented vulnerability in Android system called Stagefright was revealed by the cyber security firm \emph{Zimperium}\footnote{https://www.zimperium.com/}. Stagefright leverages the vulnerability of the media library to access users' Android devices through a simple multimedia message service (MMS) without users' awareness\footnote{http://fortune.com/2015/07/27/stagefright-android-vulnerability-text/}\footnote{http://www.forbes.com/sites/thomasbrewster/2015/07/27/android-text-attacks/}\footnote{As quoted from Zimperium chief technology officer Zuk Avraham, ``These vulnerabilities are extremely dangerous because they do not require that the victim take any action to be exploited. Unlike spear-phishing, where the victim needs to open a PDF file or a link sent by the attacker, this vulnerability can be triggered while you sleep. Before you wake up, the attacker will remove any signs of the device being compromised and you will continue your day as usual - with a trojaned phone.'' Source: http://venturebeat.com/2015/07/27/researchers-find-vulnerability-that-affects-95-of-android-devices}. As approximately 80\% of mobile devices use Android systems, nearly 1 billion devices are potential victims$^{\text{345}}$. By viewing mobile users using different operating systems as hosts with different levels of immunity to a virus, the Stagefright vulnerability behaves like the host specificity in epidemiology, as it can compromise users using Android systems.
%%%% XXX: end new writing

\section{Overview of Epidemic Models}
Here we provide an overview of classical epidemic models that have been applied to communication systems, particularly for modeling information dissemination, malware propagation, and developing the associated control methods\footnote{Due to reference count limitations only a subset of related works are introduced in this section. Interested readers can refer to \cite{Haas06,Zhang07,CPY14control,peng2014smartphone} and the references therein for more details.}.

Following terminologies from biology and epidemiology, each device in a communication system can be categorized into a few states representing its status. The main utility of such an abstraction is that one can leverage epidemic models to simplify complicated interactions among each individual and extract collective information for large-scale analysis and prediction, e.g., tracking pandemic spread patterns and predicting their outbreaks in terms of the infected population. A popular analogy is that each device is either in the \emph{Susceptible} (S), \emph{Infected} (I), or \emph{Recovered} (R) state, known as the SIR model.

For epidemic modeling of normal information dissemination dynamics, such as routing in communication networks, rumor or news spread in social networks and so on, an infected individual means he/she carries certain message (e.g., a data packet) to be delivered, a susceptible individual means he/she does not carry that message but can be potentially infected, and an recovered individual means he/she is immune to the message and hence ignores the message upon reception, e.g., in a cooperative relay-assisted network a device in the recovered state will refuse to receive or forward the packet.

For epidemic modeling of malicious codes propagation dynamics, such as privilege escalation or system vulnerability leakage, an infected individual means he/she is compromised by a malicious code and is being leveraged as a warm bed for further propagation or attack, e.g., a botnet. A susceptible individual means he/she is not comprised yet still vulnerable to the malicious code. A recovered individual means he/she is free of the threats incurred from the malicious code, e.g., securing one's devices via frequent security patch updates.

The following paragraphs introduces three basic epidemic models and relevant control techniques.

\subsection{SI Model}
SI model assumes each individual is either in the susceptible or the infected state. It can be used to estimate the reception performance of a broadcasting protocol or the dynamics of a malicious code. In \cite{CPY10}, the authors show that information dissemination in a fully mixed network of dynamic topology and opportunistic links, e.g.,  a mobile contact-based network that possesses time-varying traces due to mobility, and temporal connections due to opportunistic contacts, can be captured by an SI model. In \cite{Zou05}, the authors show that the trends of self-propagating Internet worms such as Code Red and Slammer can be successfully predicted by SI models. In \cite{Cheng11}, the authors use the SI model to formulate malware propagation in a hybrid network composed a social network and a proximal network, where a malware can leverage delocalized links (e.g., through MMS) and localized links (e.g., through Bluetooth) for propagation.

\subsection{SIS Model}
Similar to the SI model, SIS model also assumes each individual being either in the susceptible or the infected state. The difference is an SIS model allows an individual to transition from the infected state to the susceptible state. SIS models can be well mapped to the formulation of a typical two-state Markov chain where the steady-state behavior of the entire system is used for analysis. The utility of SIS models can be found in formulating recurrent network behaviors, such as the trends of receiving spam mails, or information dissemination in an evolving environment with system reconfiguration factors.
In \cite{Karyotis15}, the authors integrate the SIS model with queueing theory to study  malware propagation dynamics in a dynamic network.

\subsection{SIR Model}
SIR model is a widely used model in energy-constrained systems (e.g., a wireless sensor network) or communication systems with control capabilities over information delivery (e.g., a configurable routing protocol). An infected individual can transition from the infected state to the recovered state when certain events occur, e.g., a sensor stops to forwarding packets due to battery drain. A susceptible node can transition to the recovered state when certain mechanisms are activated, e.g., a computer is no longer vulnerable to a malicious code after installing the corresponding security patch or upgrading its operating system. In \cite{Liu09epidemic}, SIR model is used to study the vulnerability of broadcast protocols in wireless sensor networks. In \cite{Haas06,Zhang07}, SIR model is used to analyze the performance of several protocols for epidemic routing.

\subsection{Control Techniques}
One major advantage of using epidemic models for modeling dynamics of information delivery or malware propagation lies in the fact that their analytical expressions much ensemble coupled state equations appeared in control theory, which allows one to quantify a cost function of interest and evaluate the performance of a control strategy. A commonly used cost function rooted in various applications is the accumulated infected population within a time interval. For instance, in store-and-forward based routing schemes such as epidemic routing, the accumulated infected population from the time epoch when a source releases a packet to the time epoch when the packet is no longer carried by any individuals is considered as a cost function for data transmission. It can be interpreted as the system-wise buffer occupancy for data transmission since all infected devices need to keep the packet in their own buffer until the destination successfully receives the packet.

Notably, although epidemic routing enables communications in intermittently connected networks, its spreading nature inevitably induces additional system burden, especially for buffer occupancy. In \cite{Haas06}, the authors propose two strategies for controlling buffer occupancy, which we call by the \emph{global timeout scheme} and the \emph{antipacket dissemination scheme}. In the global timeout scheme, each infected individual drops the packet in its buffer when the global timer expires. In the antipacket dissemination scheme, as motivated by vaccination from immunology, upon packet reception the destination releases an antipacket as an indicator of acknowledgement (ACK) and asks every encountered individual to forward the antipacket so that infected nodes can erase the obsolete packet from its buffer,  and susceptible nodes can be prevented from receiving the already delivered packet, and hence achieving buffer occupancy reduction. 

In \cite{CPY14control}, the authors consider time-dependent control capability of SIR models in hybrid networks, where the control ability is proportional to the elapsed time, e.g., the ability to restrain malware propagation increases with the time spent in reverse-engineering its operations. An optimal control strategy based on dynamic programming is proposed for solving the optimal time epoch to implement the control strategy (analogously releasing  the antidotes) in order to balance the tradeoffs between effectiveness and consequences.

\section{Connecting the Dots: Evaluating  Transmissive Attacks via Epidemic Models}

% XXX: merge HC's improvement
Although transmissive attacks can be a serious threat to cyber security, they are often accompanied with an additional price compared with traditional attack schemes. Notably, their spreading nature and self-propagating patterns enhance the risk of exposure, and hence the attacks may be more likely to be detected. Generally speaking, while an attacker can accelerate the processes of reaching the target by compromising additional hosts, such an increased level of malicious activities becomes easily identified, thereby jeopardizing the purpose of the attack. To this end, there is a tradeoff between the probability of a successful attack and the risk of being detected due to excessive exposure.

To quantify this tradeoff between attack successfulness and risk of exposure for transmissive attacks, we propose to use epidemic models for analysis. The risk of exposure is the accumulated infected population (i.e., accumulated number of compromised hosts) from the time $0$ when the adversary launches a transmissive attack to the time $T$ when the target is comprised, or the time when the adversary decides to abort the attack, as the longer the duration of a host being compromised renders an attack more prone to be detected. The attack successfulness at a time instance $t$ is defined as the probability of successfully accessing the target between the time interval $0$ and $t$.

For further illustration, we consider the scenario where the adversary adopts the global timeout scheme for transmissive attacks as his/her control technique to reduce the risk of exposure. A global timer is set since the attacker launches a transmissive attack, and upon global timer expiration all malware residing in the compromised hosts will erase their traces via complete self-deletion, no matter the attack is successful or not, so as to allow the adversary to constrain malware propagation and alleviate exposure. 

Under the global timeout scheme an interesting question that naturally arises is: what is the optimal global timeout value $T_G$ such that the attack successfulness at time $T_G$ is no less than a certain value (e.g., 80\%) while the risk of exposure can be minimized? Partial answers to this question were given in the contexts of minimizing the system buffer occupancy while simultaneously guaranteeing end-to-end data delivery reliability between a source-destination pair for epidemic routing \cite{CPY15buffer}, where the data delivery reliability and the buffer occupancy are proven to be associated with the accumulated infected population under the SIR model \cite{Zhang07}. 

Particularly, if the mobility pattern follows a homogeneous mixing mobility assumption, such as the random waypoint model or the random direction model, a closed-form expression of optimal global timeout value is provided in \cite{CPY15buffer} that given a data delivery reliability guarantee, the optimal global timeout value that minimizes the system buffer occupancy depends on the initially infected population and the pairwise meeting rate. This suggests that if mobile botnets (i.e., several initially compromised hosts) are utilized to launch a transmissive attack, the global timer should be set smaller than that of a single seed to minimize the risk of exposure. Similarly, the global timer should decrease when the pairwise meeting rate is higher due to more frequent encounters facilitating malware propagation. Moreover, an interesting finding in \cite{CPY15buffer} is that the optimal buffer occupancy grows exponentially with the data delivery reliability. Analogously, for transmissive attacks the exponential growth rate suggests that the risk of exposure can be significantly amplified if an adversary desires higher attack successfulness. 

It is also proven in \cite{CPY15buffer} that when adopting the optimal global timer the per-user buffer occupancy does not depend on the total population for epidemic routing. This suggests that for transmissive attacks the risk of exposure for a single host can be controlled to a certain extent such that its local risk does not increase with the total host number.

\section{Experiments and Simulations}
In this section we conduct several simulations and experiments as a first step toward the analysis of transmissive attacks using epidemic models. In particular, we investigate the tradeoffs between the attack successfulness and risk of exposure by simulating global-timeout-value enabled transmissive attacks in mobile networks with two widely adopted mobility models, namely the \emph{random waypoint} (RWP) mobility model and the \emph{random direction} (RD) mobility model. We also evaluate the effect of propagation path diversity of a mobile social network on transmissive attacks based on mobile and social interaction patterns extracted from real-life datasets.
%\textcolor{blue}{In principle, the proposed attack prototype can be used for implementing transmissive attacks in such scenarios.}

\subsection{Simulation of Transmissive Attacks in Mobile Networks}

We simulate the traces of a mobile network of $N$ mobile users moving around in a wrap-around $L \times L$ square area. Any pair of users can exchange information for communication when they are within distance $r$ of each other. For RWP mobility model each user selects a destination at random and travels to the destination at a constant speed $v$. Similarly, for RD mobility model each user selects a direction at random and travels at a constant speed $v$. For both models the speed $v$  is randomly and uniformly drawn from the interval $[v_{\min},v_{\max}]$.   Initially (at time $0$) one user is compromised to launch a transmissive attack and the target is selected at random.

\begin{figure}[t]
      	\centering
      	\includegraphics[width=3.5in]{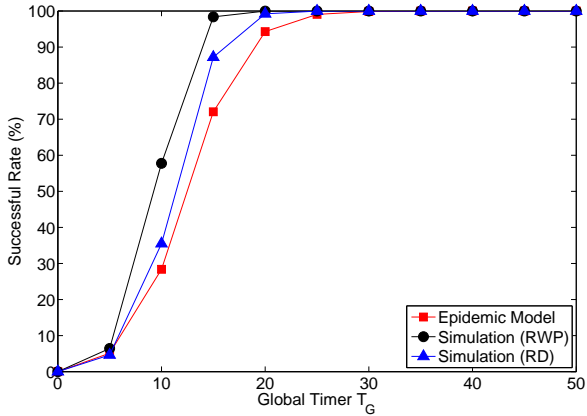}
      	\caption {Successful rate for transmissive attacks with respect to varying global timeout value $T_G$ in mobile networks simulated by RD and RWP mobility models. The system parameters are $N=100$ mobile users, $r=0.1~km$, $L=2.5352~km$,  $v_{\min}=4~km/h$, $v_{\max}=10~km/h$, and pairwise meeting rate$=0.37043$. The results are averaged over 10000 trials.}
      	\label{Fig_Sim_Suc}
      \end{figure}

\begin{figure}[t]
      	\centering
      	\includegraphics[width=3.5in]{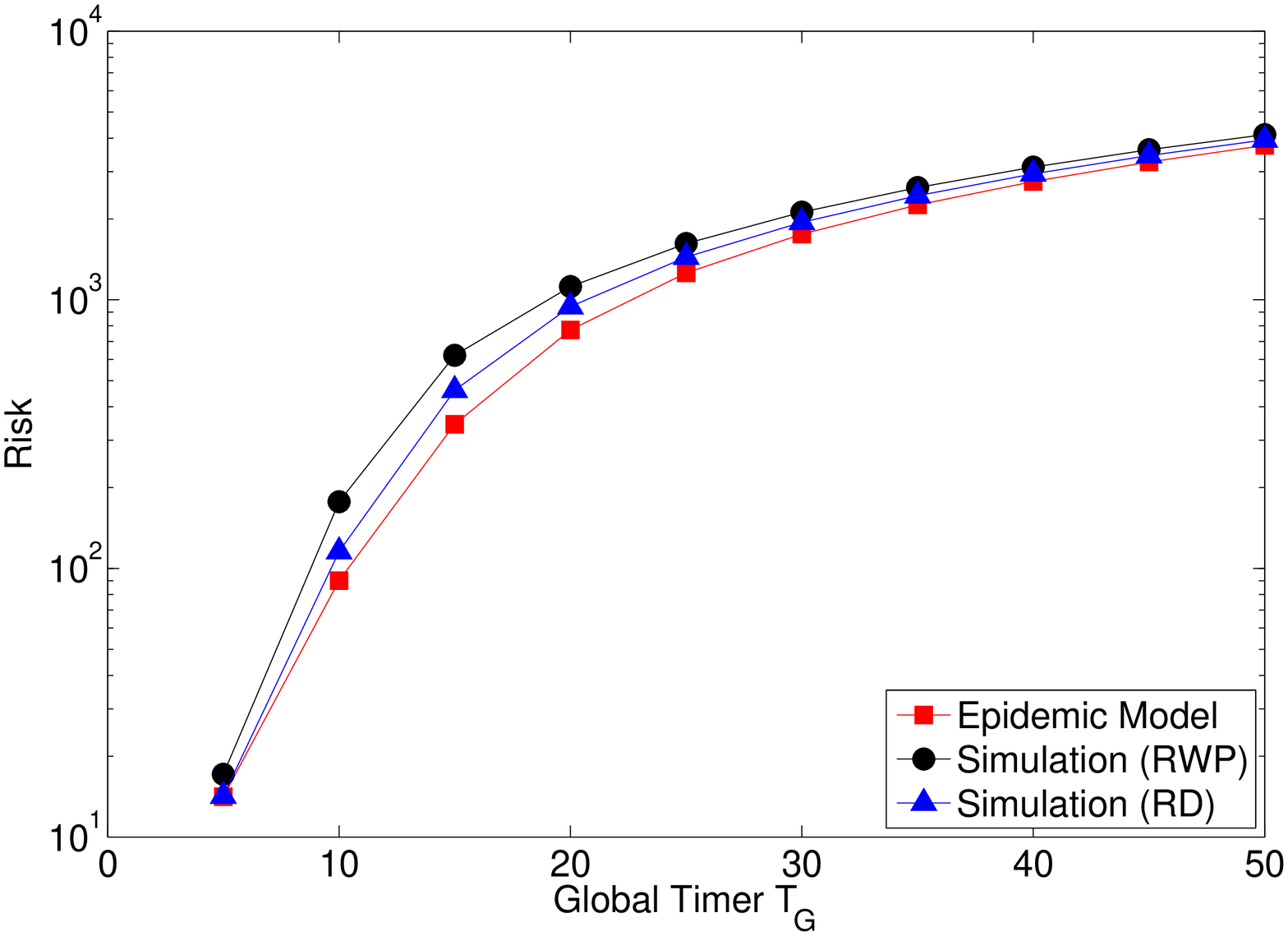}
      	\caption{Risk metric corresponding to Fig. \ref{Fig_Sim_Suc}.  Epidemic models are capable of predicting both the attack successfulness and risk of exposure. }
      	\label{Fig_Sim_Risk}
      \end{figure}

Fig. \ref{Fig_Sim_Suc} and Fig. \ref{Fig_Sim_Risk} display the attack successfulness and the risk of exposure with respect to the global timeout value $T_G$, respectively. Given $T_G$, the attack successfulness is defined as the fraction of simulated transmissive attacks that successfully approach the target prior to time $T_G$ among all trials, and the risk of exposure is defined as the accumulated compromised population divided by the total population $N$. The SIR epidemic model proposed in \cite{CPY15buffer} is used for performance comparison. The successful rate is the probability of infecting a particular host, and the risk of exposure is evaluated using the accumulated infected population.

It can be observed that the global timeout value $T_G$ indeed governs the performance of both attack successfulness and risk of exposure. The simulation results also validate the tradeoffs between these two metrics as the enhancement in attack vulnerability often leads to the increase in risk of exposure, and vice versa. For example, to enhance the attack successfulness from 30 \% ($T_G=10$) to 90 \% ($T_G=20$), the risk of exposure needs to be amplified by 10 times. Notably, the predicted results from the epidemic model can successfully capture the trends of these two metrics. An immediate utility is that an adversary can use the epidemic model to determine the optimal global timeout value that guarantees the attack successfulness while simultaneously minimizing the risk of exposure, e.g., selecting $T_G=25$ such that the attack successfulness is no less than 95 \%. Moreover, a defender can also utilize the epidemic model to evaluate a system's vulnerability without conducting time-consuming simulations.

\subsection{Experiment of Transmissive Attacks in Mobile Social Networks}

      \begin{figure}
      	\centering
      	\includegraphics[width=3.5in]{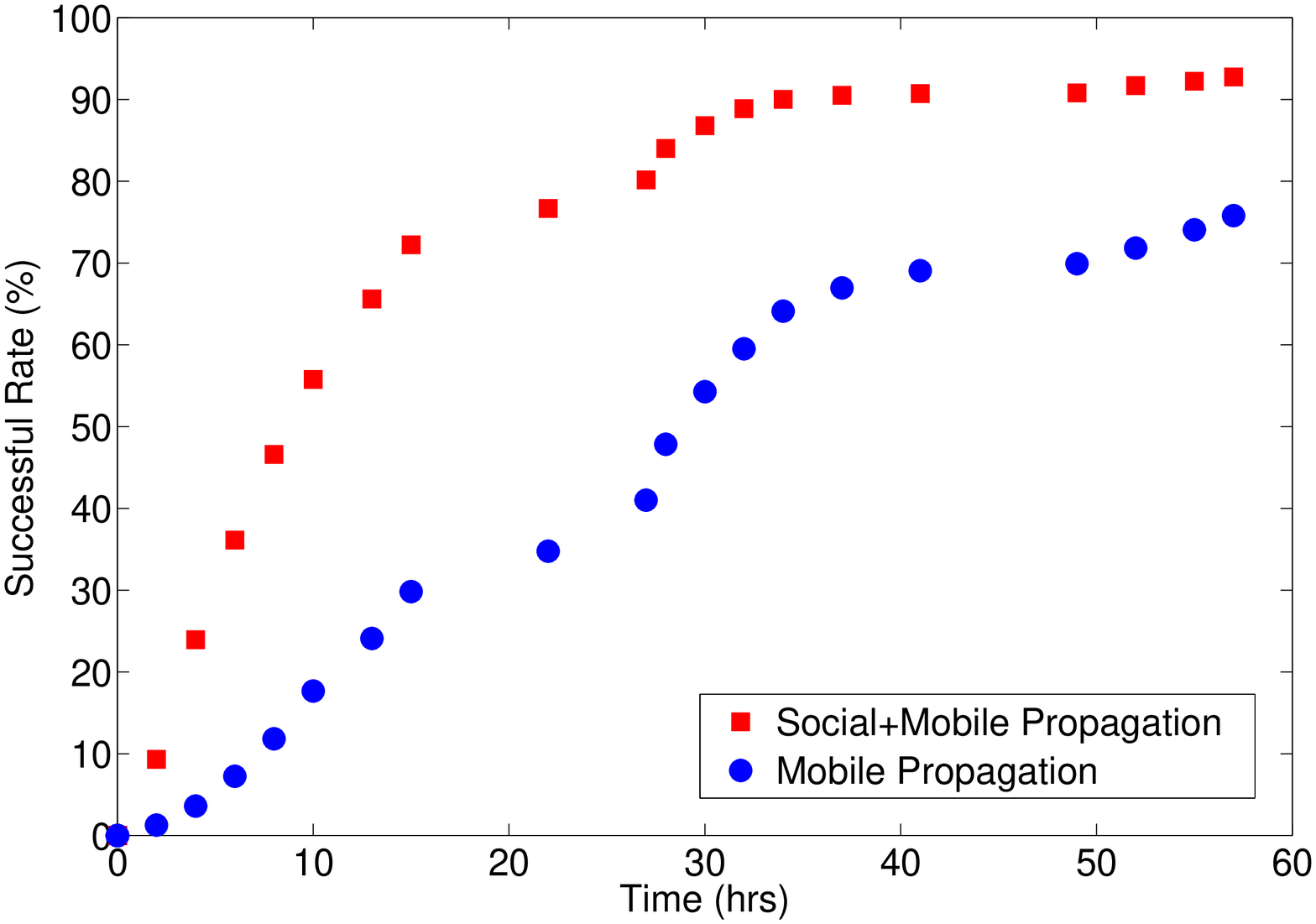}
      	\caption{Successful rate for transmissive attacks in a mobile social network: exploiting both social and mobile propagation paths can significantly improve the possibility of approaching the target. The propagation parameters $p_s=p_\ell=0.05$, and the results are averaged over $10000$ trials.}
      	\label{Fig_Dataset_Suc}
      \end{figure}
      
      \begin{figure}
      	\centering
      	\includegraphics[width=3.5in]{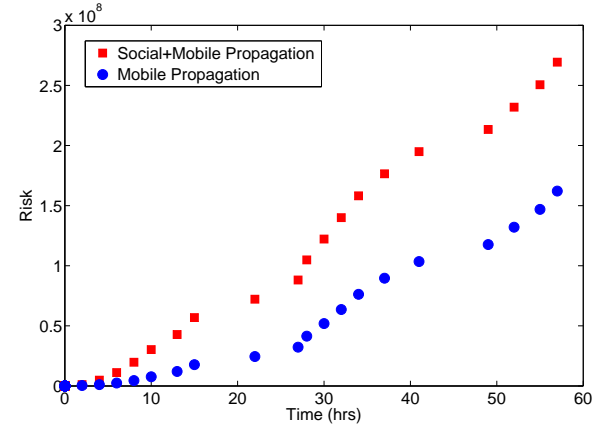}
      	\caption{Risk metric corresponding to Fig. \ref{Fig_Dataset_Suc}: the accumulated infected population with respect to time. The results show clear tradeoffs between attack successfulness and risk of exposure.}
      	\label{Fig_Dataset_Risk}
      \end{figure}

To investigate the impact of propagation path diversity on transmissive attacks, we use the CRAWDAD mobile-social interaction traces\footnote{CRAWDAD dataset thlab/sigcomm2009 (v. 2012-07-15). Available at http://crawdad.org/thlab/sigcomm2009/20120715} to simulate a transmissive attack. The purpose of this experiment is to study the consequences of transmissive attacks that are capable of propagating through social contacts (e.g., via MMS) or proximity contacts (e.g., via Bluetooth). In such a mobile social network the malware can propagate from one compromised user to another user with probability of success $p_s$ via the social propagation path if these two users are social contacts, i.e., there is an edge between these two users in the corresponding social graph. Similarly,  the malware can propagate from one compromised user to another user with probability of success $p_{\ell}$ via the proximity propagation path if these two users are within a physical contact distance.

Fig. \ref{Fig_Dataset_Suc} and Fig. \ref{Fig_Dataset_Risk} display the attack successfulness and risk of exposure for transmissive attacks in the mobile social network, respectively. 
%The initial and final time instances are set to be the same as the time records in the contact dataset.
It can be observed that the inclusion of social propagation paths can significantly enhance the attack successfulness. For example, after 30 hours since launching a transmissive attack, the attack successfulness of utilizing both social and mobile propagation paths can be doubled compared with the attack successfulness of only exploiting mobile propagation paths. However, the induced risk metric is also amplified as shown in Fig. \ref{Fig_Dataset_Risk}. 

Additional experiments of different parameter configurations show similar trends in attack successfulness and risk of exposure, which are discussed in the supplementary file. These results suggest that propagation path diversity can facilitate transmissive attacks at the price of potentially amplified exposure. In addition, how current epidemic models can be improved to model transmissive attacks in such a heterogeneous network is an active research area.

\section{Some Ongoing Challenges and Open Research Questions}
Here we discuss several ongoing challenges and open research questions related to transmissive attacks.

\begin{itemize}
	\item \textbf{Lateral movement detection and prevention.}~\\
%	Recently lateral movement attacks have drawn many attention in security society, as researches gradually reach a consensus that a cyber system needs to remain operation even when it is potentially under attacks. 
Unlike disruption attacks (e.g., denial of service) that often cause distinguishable anomalous activities, lateral movement attacks (e.g., privilege escalation that insidiously acquires user credentials) are difficult to detect. 
	Transmissive attacks fall into one category of lateral movement attacks due to their stealthy transmissive nature. 
If detecting lateral movement is implausible, one may shift attention to designing a resilient cyber system that can constrain the damage induced by such attacks.

	\item \textbf{Transmissive attacks in network of networks (NoNs).}~\\
	Network of networks (NoNs) is an intuitive explanation of modern communication systems with intrinsic layered structures and heterogeneous networks. The layers of the Internet architecture can be operated by different protocols,
	and a device can have multiple communication resources (e.g., cellular, WiFi and Bluetooth modules).
	 As horizontal malware propagations within a single layer/system can be straightforward by leveraging similar vulnerabilities, vertical malware propagations traversing different layers/systems can be more difficult due to lack of common vulnerabilities or implementation of additional security rules. In terms of bio-inspired attacks, transmissive attacks that are self-evolving and adaptive to the NoN environment can be a vital threat.

	\item \textbf{Data-driven inference for attack and defense.}~\\
	In a data-rich era, our cyber footage is everywhere and easy to track. Both attackers and defenders should make use of available data collected from different sources to infer vulnerabilities in a system.  Notably, modern technology enables an adversary to optimize his/her attack strategy based on the inference results from the collected data prior to launching a transmissive attack, known as the inference attacks. For instance, personal trace information such as GPS signals or locations revealed by online social networking activities can be directly observed or indirectly inferred from user-centric data.

	\item \textbf{Evolutionary resilience of dynamic systems.}~\\
	In many cases the underlying communication system where a transmissive attack takes place is an ever-changing system due to variations in time, traffic flows, evolution of communication technology, and so on. Therefore, a general notion of resilience for such a dynamic system is necessary to quantify network stability that can vary with time, which we call as \emph{evolutionary resilience}. Notably, biology models such as ecological systems, predator-prey models, and evolutionary game theory that target at evolutionary stability in time-varying coupled systems may be well mapped to analyze transmissive attacks in dynamic systems.	
\end{itemize}

\section{Conclusion}
This paper introduces an emerging attack pattern called transmissive attack that leverages diverse communication paths to approach the target and accomplish its task. Inspired by biology, we provide an overview of commonly used epidemic models for communication systems and connect the dots between transmissive attacks and epidemic models. We perform simulations via two widely used mobility models and conduct experiments in mobile social networks to demonstrate the utility of epidemic models for assessing attack successfulness and risk of exposure, and we also discuss some ongoing research challenges and open research questions related to transmissive attacks.

%\begin{figure*}[t]
%	\centering
%	\includegraphics[width=5in]{fig_matrix}
%	\caption{Illustration of crowdsourced data.}
%	\label{fig_matrix}
%\end{figure*}

%\cite{Adamic05}

\clearpage
\section{Supplementary File}
This seciton is a supplement that illustrates horizontal and vertical propagation involved in transmissive attacks and
includes additional experimental results of simulating transmissive attacks with the CRAWDAD mobile-social interaction traces\footnote{CRAWDAD dataset thlab/sigcomm2009 (v. 2012-07-15). Available at http://crawdad.org/thlab/sigcomm2009/20120715}. 

\begin{figure}
	\centering
	\includegraphics[width=3.3in]{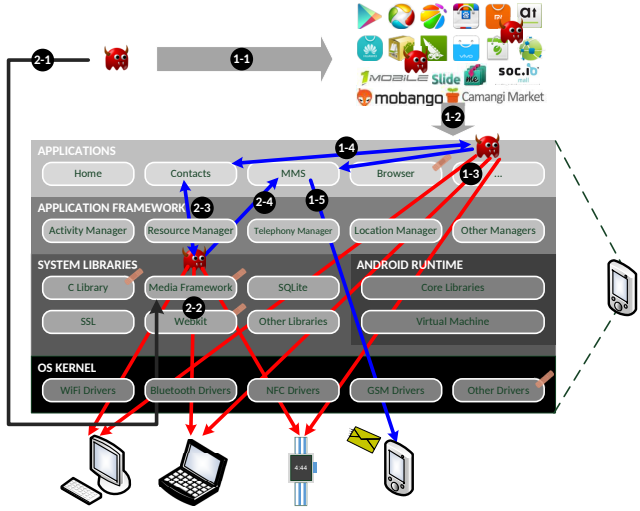}
	\caption{The block diagram of common components in an Android system and how they are involved in the propagation of transmissive attacks. An attacker can deploy malicious codes to application market places and wait for unaware users to install the deployed codes passively (paths 1-1 and 1-2), or compromise a device by using remote vulnerabilities and inject executables directly into the device (path 2-1). Once malicious codes are in operation, they can discover nearby network links (paths 1-3 and 2-2) and social links (paths 1-4 and 2-3), and then attempt to launch transmissive attacks at either the network level  (paths 1-3 and 2-2) or the application level (paths 1-5 and 2-4). }
	%\vspace*{4in}
	\label{Fig_attack_flow}
\end{figure}

We first show how attackers compromise a mobile device and propagate malicious codes to nearby devices in Fig. \ref{Fig_attack_flow}. An attacker usually has two choices to gain access to a device. One is to deploy malicious codes to application market places and wait for unaware users to install the deployed codes passively (paths 1-1 and 1-2). The other is to compromise a device by using remote vulnerabilities and inject executables directly into the device (path 2-1). Once malicious codes are in operation, they can discover nearby network links (paths 1-3 and 2-2) and social links (paths 1-4 and 2-3), and then attempt to launch transmissive attacks at either the network level  (paths 1-3 and 2-2) or the application level (paths 1-5 and 2-4). Remote vulnerabilities are important for successful transmissive attacks, since mobile and wearable devices may be not accessible via Internet. Therefore, attackers have to identify possible remote vulnerabilities in the system. For example, the recent vulnerabilities identified in Stagefright library, which is a part of Media Framework in Fig. \ref{Fig_attack_flow}, is one good option to launch transmissive attacks: it is a built-in component that is associated with many services and applications.

	\begin{figure}[t]
		\centering
		\begin{subfigure}[b]{0.45\linewidth}
			\includegraphics[width=\textwidth]{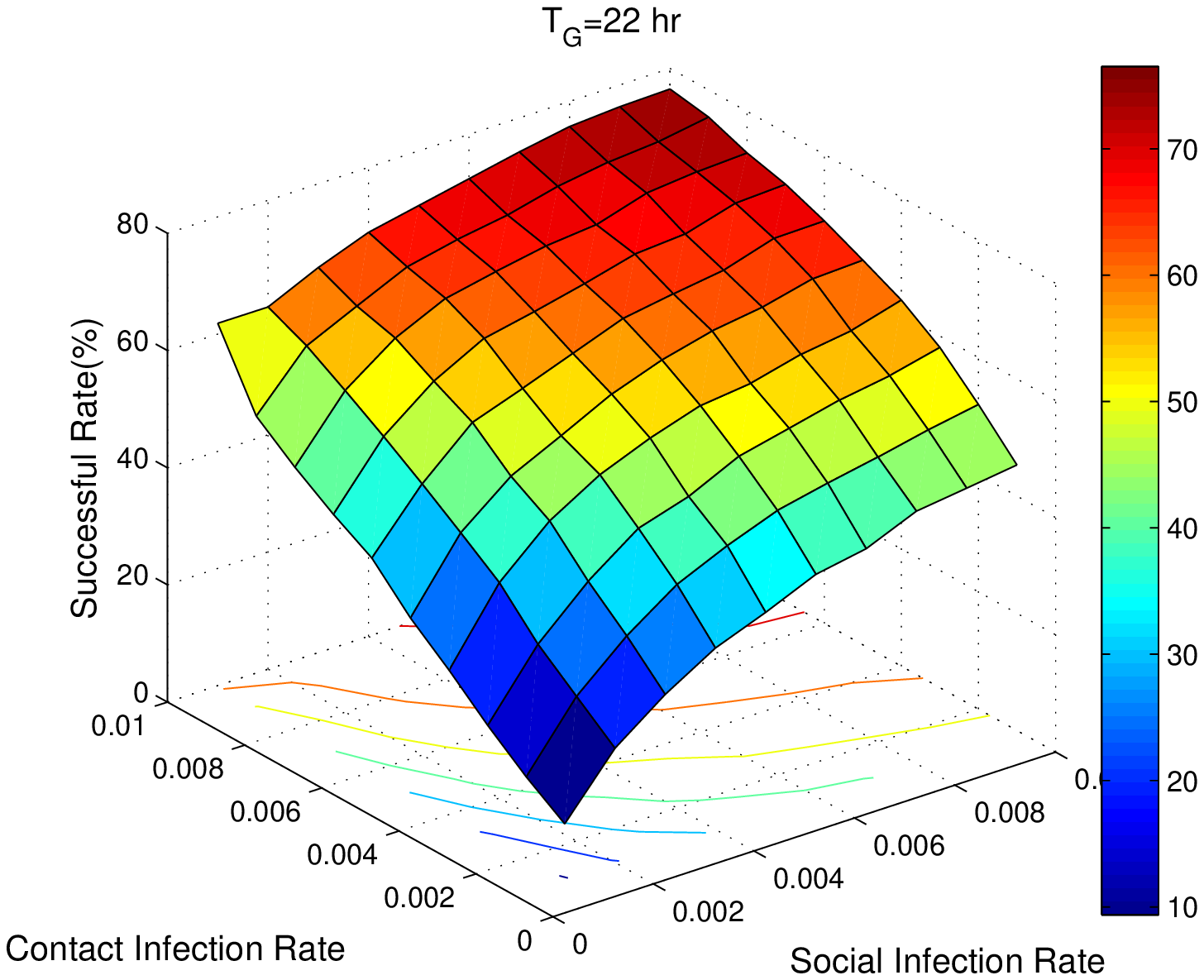}
			\caption{$T_G=22$ hours.}
			%			\label{}
		\end{subfigure}%
		%		\hspace{3.8cm}
		\centering
		\begin{subfigure}[b]{0.45\linewidth}
			\includegraphics[width=\textwidth]{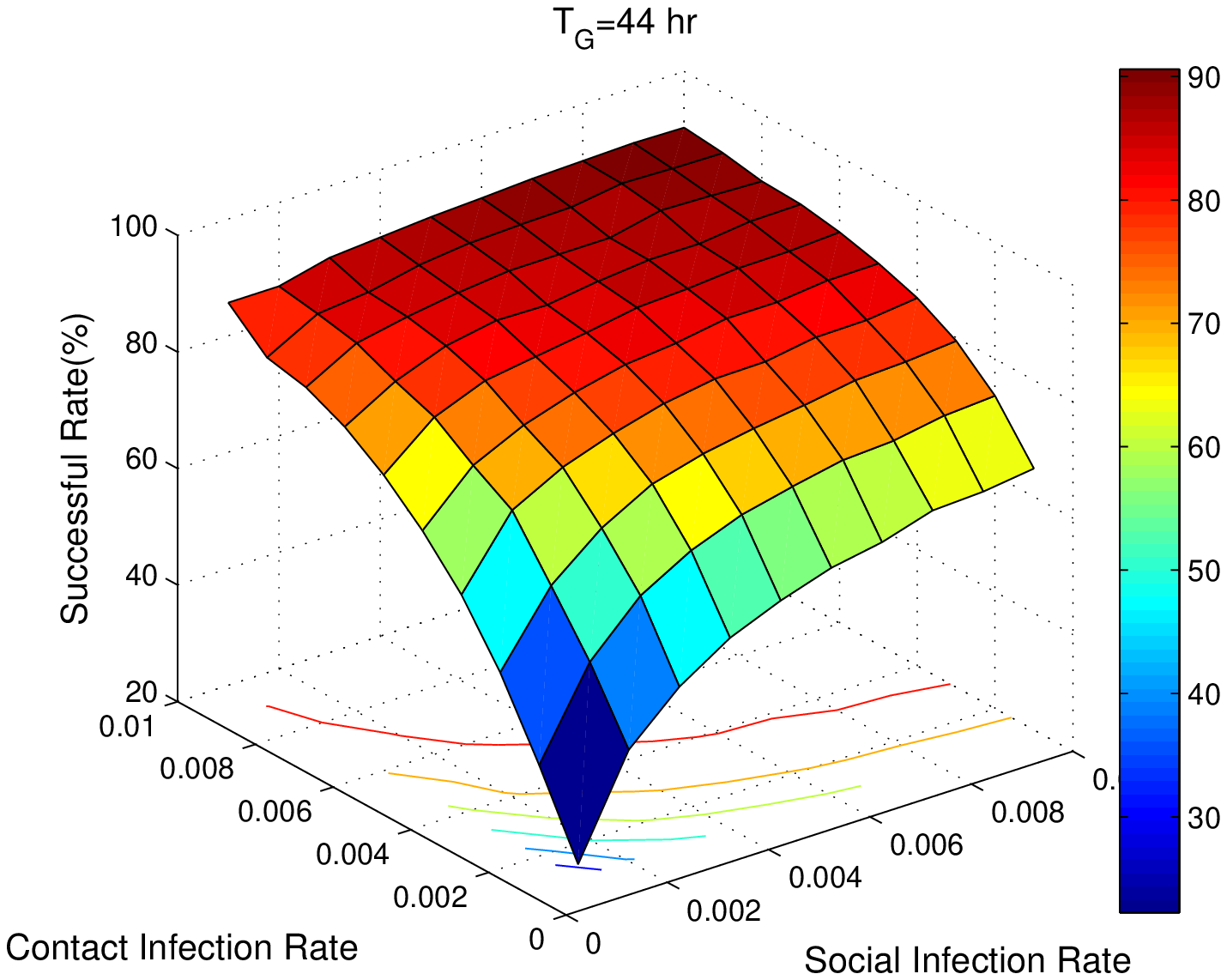}
			\caption{$T_G=44$ hours.}
			%			\label{}
		\end{subfigure}
		\centering
		\begin{subfigure}[b]{0.45\linewidth}
			\includegraphics[width=\textwidth]{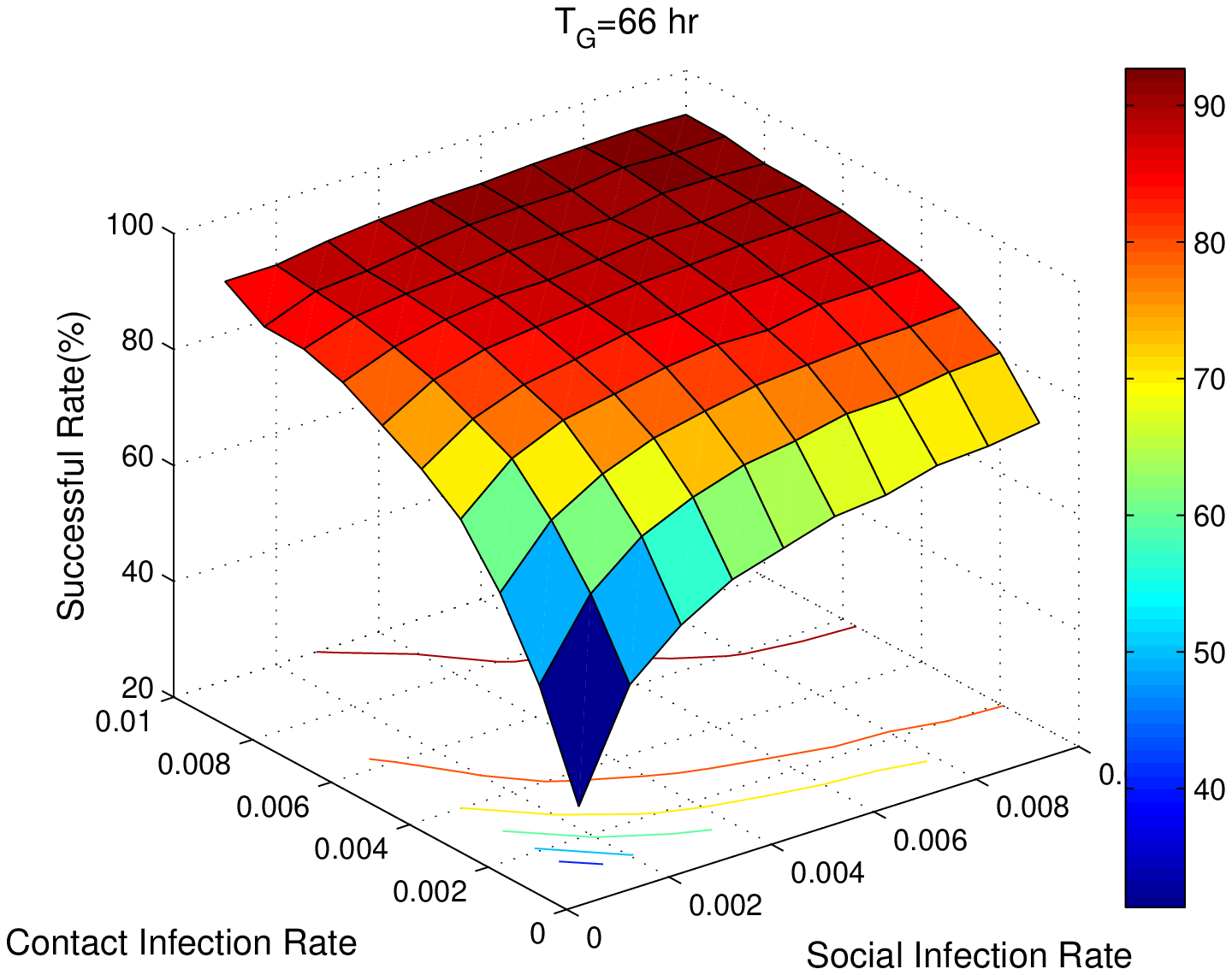}
			\caption{$T_G=66$ hours.}
			%			\label{}
		\end{subfigure}
		\centering
		\begin{subfigure}[b]{0.45\linewidth}
			\includegraphics[width=\textwidth]{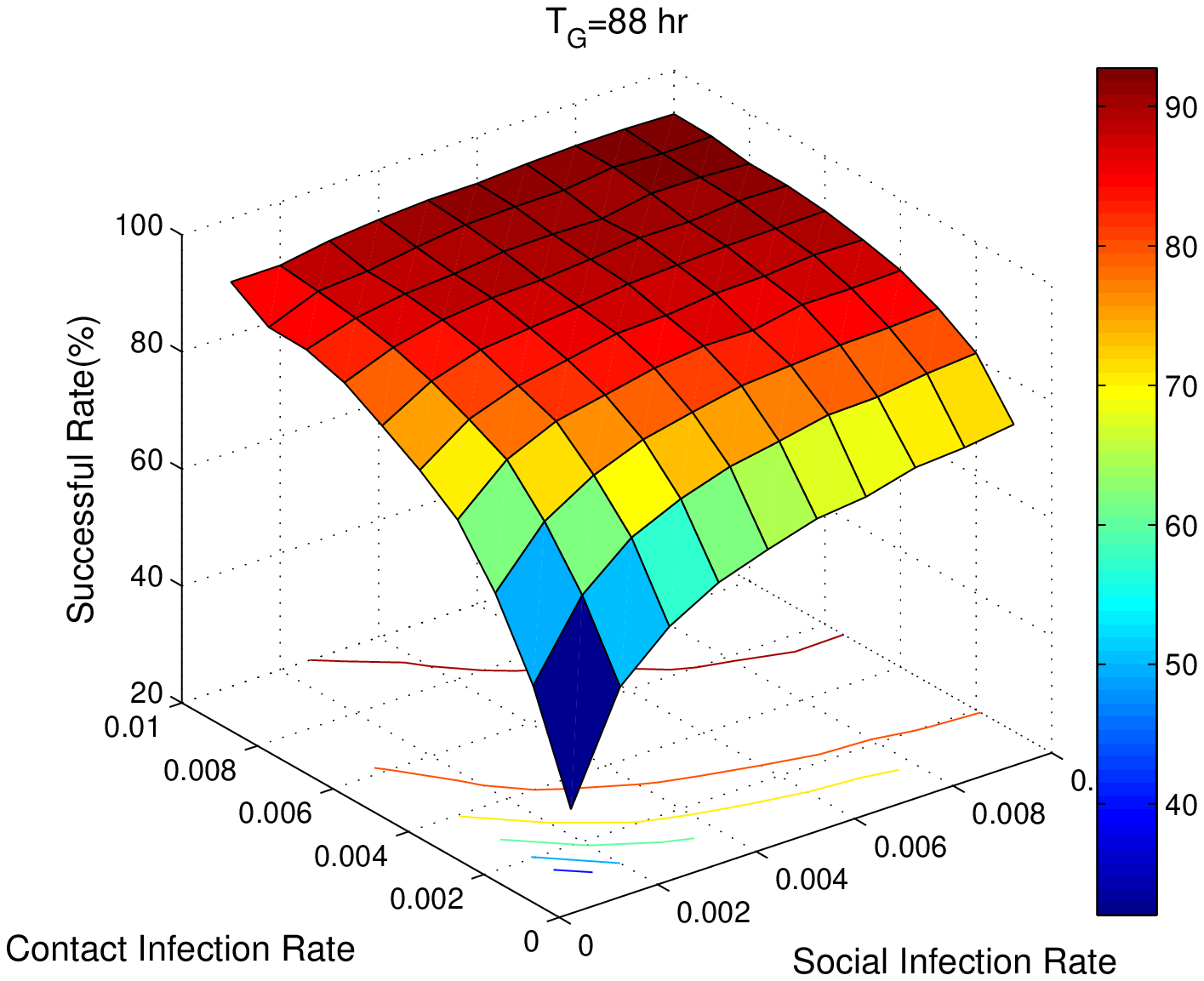}
			\caption{$T_G=88$ hours.}
			%			\label{}
		\end{subfigure}	
		\caption{Successful rate for transmissive attacks under different configurations. The results are averaged over $10000$ trials. For different time instances, it is observed that the increase in $p_s$ or $p_\ell$ can facilitate transmissive attacks.}
		\label{Fig_attack_3D}	
	\end{figure}

	\begin{figure}[t]
		\centering
		\begin{subfigure}[b]{0.45\linewidth}
			\includegraphics[width=\textwidth]{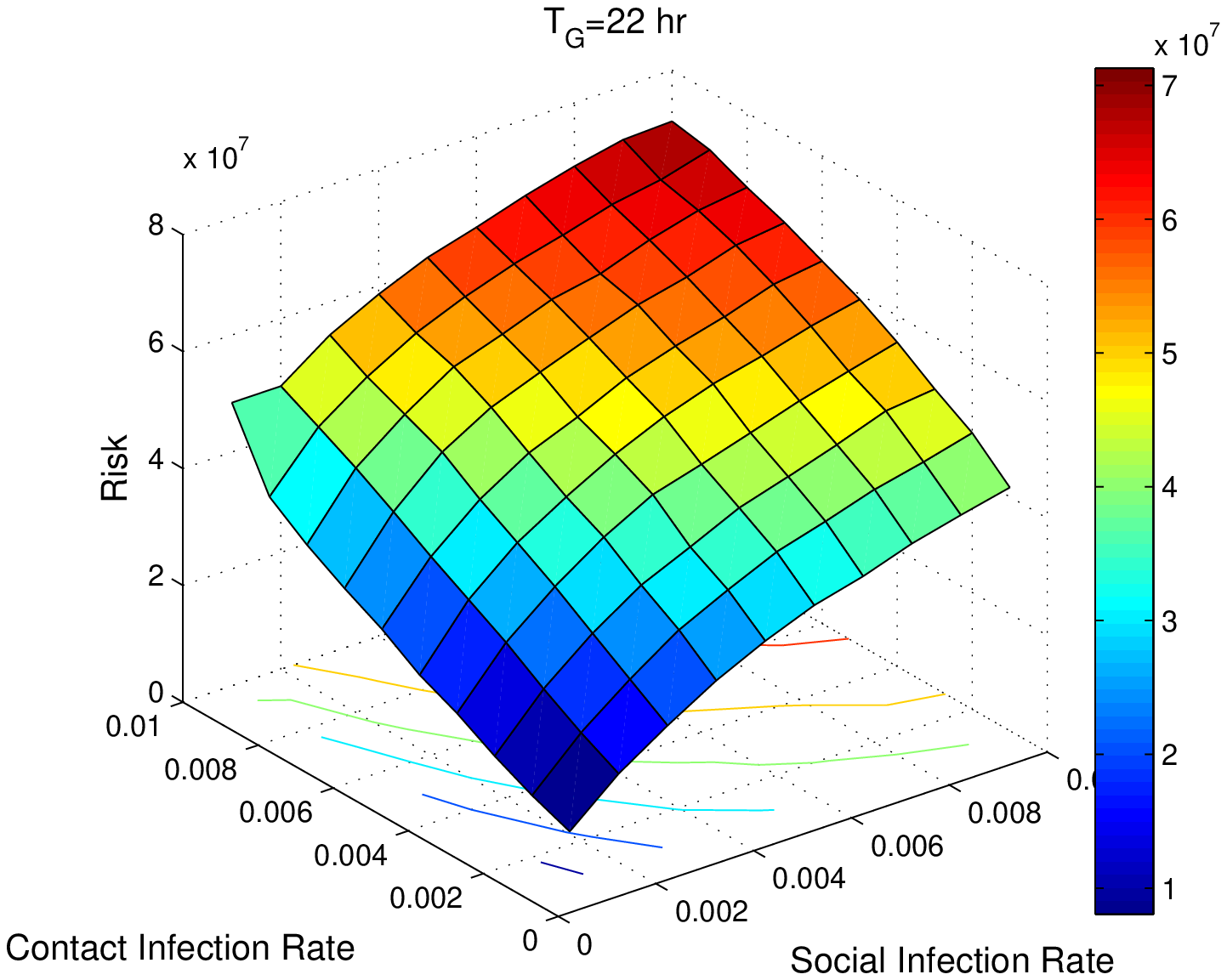}
			\caption{$T_G=22$ hours.}
			%			\label{}
		\end{subfigure}%
		%		\hspace{3.8cm}
		\centering
		\begin{subfigure}[b]{0.45\linewidth}
			\includegraphics[width=\textwidth]{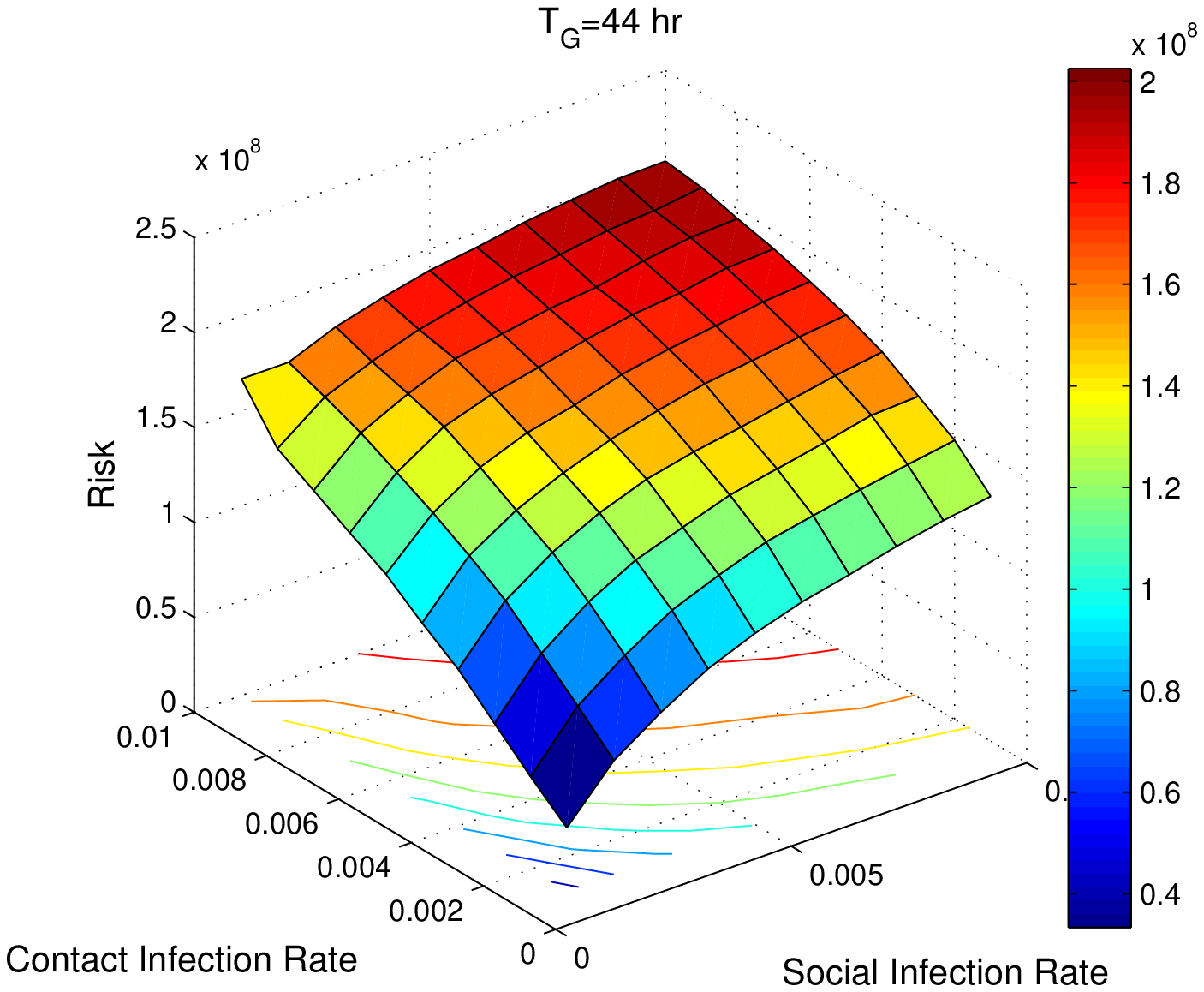}
			\caption{$T_G=44$ hours.}
			%			\label{}
		\end{subfigure}
		\centering
		\begin{subfigure}[b]{0.45\linewidth}
			\includegraphics[width=\textwidth]{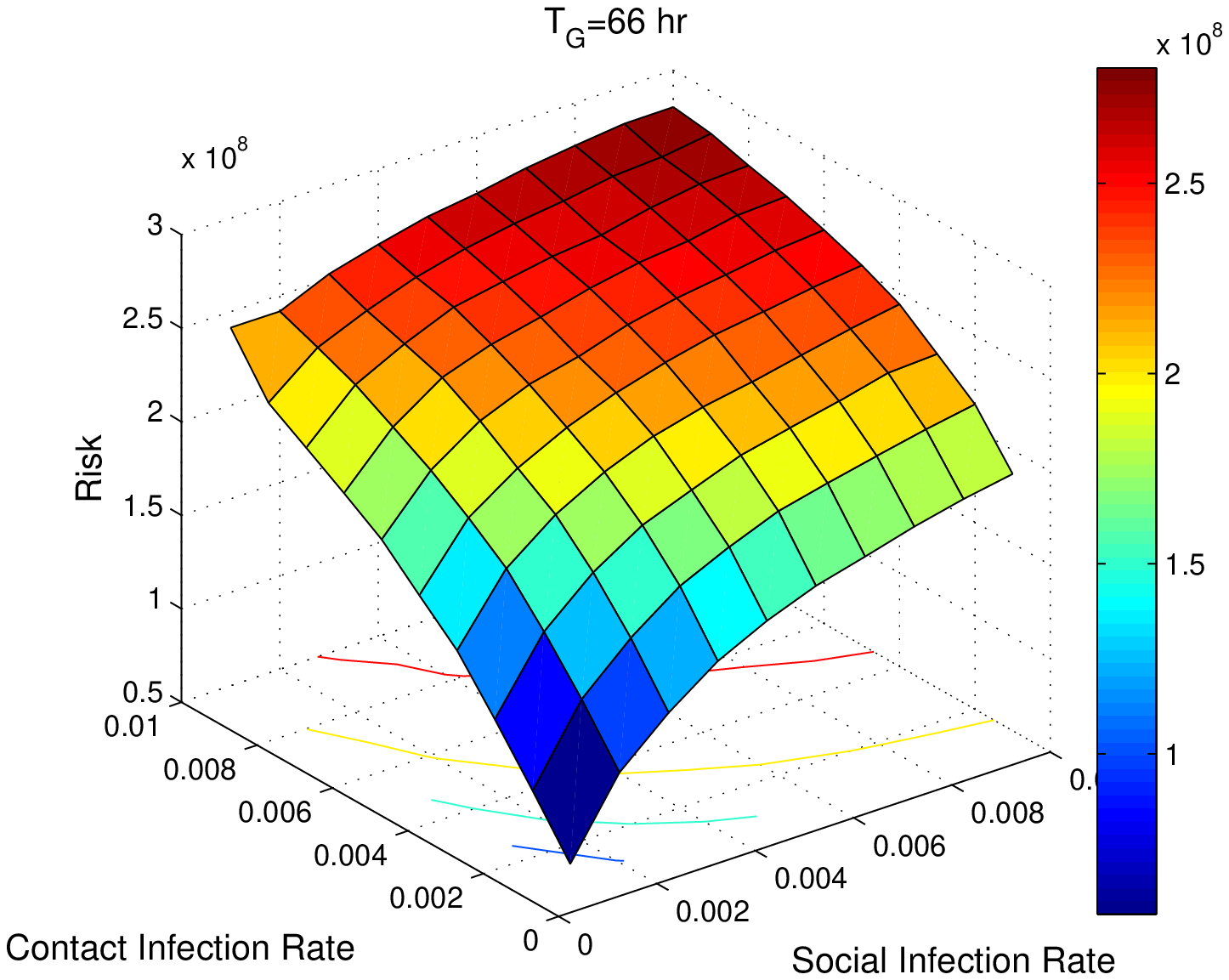}
			\caption{$T_G=66$ hours.}
			%			\label{}
		\end{subfigure}
		\centering
		\begin{subfigure}[b]{0.45\linewidth}
			\includegraphics[width=\textwidth]{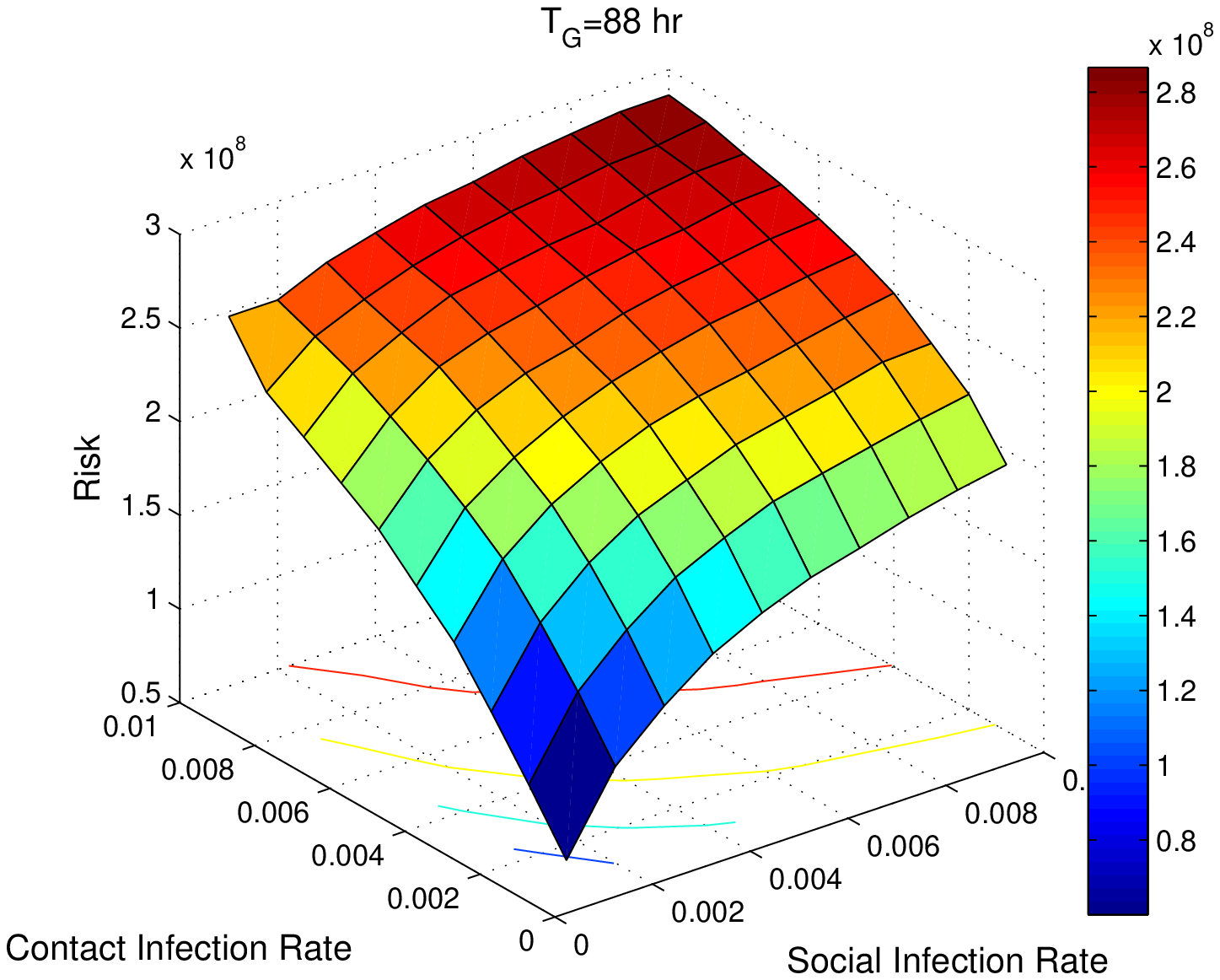}
			\caption{$T_G=88$ hours.}
			%			\label{}
		\end{subfigure}	
		\caption{Risk metric for transmissive attacks under different configurations. The results are averaged over $10000$ trials. For different time instances, it is observed that the increase in $p_s$ or $p_\ell$ also leads to higher risk.}	
	\label{Fig_risk_3D}		
 	\end{figure}

We investigate the attack successful rates and the corresponding risk metrics under different configurations, namely the social and proximal infection rates $p_s$ and $p_\ell$, for performance comparison. Fig. \ref{Fig_attack_3D} and Fig. \ref{Fig_risk_3D} show the attack successfulness and the corresponding risk of exposure with varying configurations at four different time instances, respectively. It is observed that both the attack successfulness and the risk metric increase with the configuration parameters $p_{s}$ and $p_{\ell}$. These results indicate nontrivial tradeoffs between the attack successfulness and the risk metric. 

If the attacker has the freedom of manipulating $p_s$ and $p_\ell$, then assigning maximal values to these two parameters is usually not the best strategy, since doing so incurs huge risk, and thereby the attacks can be in vain due to high exposure. Instead, the optimal strategy of an attacker is to first identify the risk the attacker is willing to take, and then find a set of configuration parameters that maximize the successful rate, which can be casted as a configuration parameter optimization problem with risk constraints.

% \bibliographystyle{IEEEtran}
% \bibliography{IEEEabrv,Comm_sup}

\end{document}